\documentstyle[12pt]{article}
\def \be {\begin{equation}}
\def \ee {\end{equation}}
\advance\oddsidemargin by -1.2cm
\advance\evensidemargin by -1.2cm
\begin{document}
\baselineskip 0.33in
\parindent=1cm
\large

\vspace*{3cm}

\begin{center}
{\LARGE \bf  Spin-Wave Theory of the Spiral  Phase of the t-J Model}

\

\

 {\it N.I.Karchev \\ Department of Physics, University of Sofia, 1126 Sofia,
Bulgaria

T.S.Hristov \\
Space Research Institute, Bulgarian Academy of Sciences, 1000 Sofia, Bulgaria}
\end{center}

\

\

{\bf Abstract}

\

A graded Holstein-Primakoff realization of the SU(2/1) algebra is proposed. A
spin-wave theory with a condition that the sublattice magnetization is zero is
discussed. It is shown that the generalized  spin-wave theory is appropriate to
investigate the long-range spiral  (incommensurate) phase
of the $t-J$ model at $T=0$. The spin-spin correlator is
calculated.

\

\

\

PACS Numbers: 75.10.Jm, 74.20.-z, 74.65.+n

\newpage
\baselineskip 0.3in

{\bf I. Introduction}

Among the many electronic models which are being currently studied in the
context of high-T$_c$ superconductivity, the two dimensional $t-J$ model is the
simplest one and it is rather general to describe the basic physics of the new
superconductor$^1$. It is an effective model for the large $U$ Hubbard model
with $J = {4t^2\over U}$, and there are arguments that one can derive it
from a multiband Hubbard model$^{2,3}$.

At half-filling the $t-J$ model reduces to the Heisenberg model. This is a
welcome result, since, the methods developed for the Heisenberg model can be
extended and applied to investigate the properties of the $t-J$ model, atleast
at nearly half-filling. Schwinger boson slave-fermion mean field theory was
used and an antiferromagnetic ordered Ne\'el state at half-filling was
found$^4$. This state evolves into long-range spiral (incommensurate)
antiferromagnetic states at large doping.  Quantitatively, this is described by
the dependence of the spirality angle $Q$ on the doping$^5$.  Originally, the
spiral phase
of the $t-J$ model was obtained by Shraiman and Siggia who used a somewhat
different approach$^6$.

Recently, Takahashi has formulated a modified spin-wave theory of Heisenberg
(anti)ferromagnets$^{7,8}$. He has supplemented the usual spin-wave theory with
the constraint that the magnetization of each site is zero. This enforces the
condition that the total number, on average, of spin waves per site is $S$, and
that the sublattice rotational symmetry is not broken. In one dimension the
modified spin wave theory yields excellent agreement with Bethe ansatz results
for spin $S = 1/2$. Takahashi's results are in quantitative agreement with the
Schwinger boson mean field theory$^4$ and with the renormalization group
theory$^9$.

In this paper we formulate, along the same line, a generalized spin wave theory
appropriate to investigate the spiral phase of $t-J$ model. Some results of
Ref. 10 are used.

In Section II the canonical bose- and fermi- operators are used to realize the
SU(2/1) algebra. The graded Holstein-Primakoff re\-pre\-sen\-ta\-tion is
written in a
local frame. The effective hamiltonian is obtained in leading order of
$S^{-1}$, as a function of the angles $\theta_i$ which determine the local
frame. To preserve the sublattice rotational symmetry we impose two additional
conditions. To enforce them, we introduce two terms in the hamiltonian with
chemical potentials $\lambda$ and $\zeta$. For the special values of $\theta_i
= k \pi$, $\zeta$ is equal to zero and we reobtain the Takahashi's spin-wave
theory. We express the angles $\theta_i$ by the spirality angle $Q$. To
determine
the last one, we calculate the energy of the system per site. The physical
value of $Q$ is that which minimizes the energy. It depends on the doping.

In Section III we calculate the large distance asymptotic of the spin-spin
correlation function. At $T = 0$ we obtain long-range correlation which is
characterized by the angle $Q$ and the quantity $m$ which is related to the
spin-wave bose-condensate. At finite temperature, the correlation function
exponentially falls when the separation between the two sites is large. At
small doping the spin-spin correlator looks like in the theory of Heisenberg
antiferromagnet. When the doping is large enough it looks like in the theory of
Heisenberg ferromagnet.

Section IV is devoted to the concluding remarks.
\ \\

{\bf II. Graded spin-wave theory}

 The $t-J$ model is defined by the Hamiltonian
\be
h = -t \sum_{<i,j>} \left[ c^+_{i\sigma} c_{j\sigma} + h.c.\right] + J
\sum_{<i,j>} \left( \vec{S}_i \cdot \vec{S}_j - {1\over 4} n_i n_j \right) -
\mu \sum_{i} c^+_{i\sigma} c_{j\sigma}
\ee
where $c^+_{i\sigma} (c_{j\sigma})$ are the fermi operators of the electron
($\sigma = 1,2$) on site $i(j)$ of a $2D$ square lattice which act on states
with no double occupancy on a lattice site $i(j)$, $\vec{S}_i$ are spin
operators, and $\mu$ is the chemical potential. By $<i,j>$ we denote the sum
over the nearest neighbors.

The eight operators $c_{i\sigma}$, $c^+_{i\sigma}$, $S^{\pm}_{i}$, $S^3_i$ and
$n_i$ form a basis of the SU(2/1) graded algebra. It is most convenient to
rewrite them in terms of the Hubbard operators $X^{ab}$ ($a,b = 1,2,3$)
\be
\begin{array}{lll}
c_{i\sigma} = X^{3\sigma}_i & c^+_{i\sigma} = X^{\sigma 3}_i & \sigma = 1,2 \\
\ \\
n_i = X^{11}_i + X^{22}_i & S^3_i = {1\over 2} \left( X^{11}_i - X^{22}_i
\right) \\
\ \\
S^+_i = X^{12}_i & S^-_i = X^{21}_i
\end{array}
\ee
They satisfy the following graded commutation rules
\be
\left[X^{ab}_i, X^{cd}_j\right]_{\pm} = \delta_{ij} \left(\delta^{bc} X^{ad}_i
\pm \delta^{ad} X^{cb}_i \right)
\ee

The graded algebra SU(2/1) can be realized using a canonical bose $a_i, a^+_i$
and fermi $\psi_i, \psi^+_i$ operators
\be
\left[a_i, a^+_j\right]_- = \delta_{ij}; \ \ \ \ \  \left[\psi_i,
\psi_j^+\right]_+ = \delta_{ij}
\ee
\ \\
$$
c_{i1} = \psi^+_i \left[ \left(2S - a^+_i a_i - \psi^+_i \psi_i \right)^{1/2}
\cos {\theta_i\over 2} + a_i \sin {\theta_i\over 2} \right]
$$
$$
c^+_{i1} =  \left[ \left(2S - a^+_i a_i - \psi^+_i \psi_i \right)^{1/2}
\cos {\theta_i\over 2} + a^+_i \sin {\theta_i\over 2} \right] \psi_i
$$
$$
c_{i2} =  \psi^+_i \left[ - \left(2S - a^+_i a_i - \psi^+_i \psi_i
\right)^{1/2} \sin {\theta_i\over 2} + a_i \cos {\theta_i\over 2} \right]
$$
$$
c^+_{i2} =  \left[ - \left(2S - a^+_i a_i - \psi^+_i \psi_i \right)^{1/2}
\sin {\theta_i\over 2} + a^+_i \cos {\theta_i\over 2} \right] \psi_i
$$
\be
n_i = 2S - \psi^+_i \psi_i
\ee
$$
S^3_i = {1\over 2} \sin \theta_i \left[ \left(2S - a^+_i a_i -
\psi^+_i \psi_i
\right)^{1/2} a_i + a^+_i \left(2S - a^+_i a_i - \psi^+_i \psi_i \right)^{1/2}
\right]
$$
$$
+ \cos \theta_i \left(S - a^+_i a_i - {1\over 2} \psi^+_i \psi_i
\right)
$$
$$
S^+_i =  \cos^2 {\theta_i\over 2}  \left(2S - a^+_i a_i - \psi^+_i \psi_i
\right)^{1/2} a_i - \sin^2{\theta_i\over 2}
a^+_i \left(2S - a^+_i a_i - \psi^+_i \psi_i \right)^{1/2}
$$
$$
-  \sin  \theta_i \left(S - a^+_i a_i - {1\over 2} \psi^+_i \psi_i
\right)
$$
$$
S^-_i =  \cos^2 {\theta_i\over 2}  a^+_i \left(2S - a^+_i a_i - \psi^+_i \psi_i
\right)^{1/2}  - \sin^2{\theta_i\over 2}
\left(2S - a^+_i a_i - \psi^+_i \psi_i \right)^{1/2} a_i
$$
$$
-  \sin  \theta_i \left(S - a^+_i a_i - {1\over 2} \psi^+_i \psi_i
\right)
$$
where $\theta_i$ are arbitrary angles which run over intervals with length
$\pi$. When $\theta_i = 0, \pi , 2\pi ...$ the Eqs. (5) reduce to the graded
Holstein-Primakoff re\-pre\-sen\-ta\-tion$^{11}$. It is easy to prove that so
defined
operators satisfy the algebra Eqs.(2,3). In particular, the spin operators
satisfy the SU(2) algebra and $S$ parametrizes the representation of the
SU(2/1) algebra. At half-filling $S$ is exactly the spin of the system. The
physical relevant case is $S = {1\over 2}$.

In spin-wave approximation one must replace  $\left(2S - a^+_i a_i - \psi^+_i
\psi_i \right)^{1/2}$ by \\  $(2S)^{1/2} \left[ 1 - {1\over {4S}}  \left(a^+_i
a_i + \psi^+_i \psi_i \right) \right]$ in Eqs. (5). Then, in leading order of
$S^{-1}$, the Hamiltonian Eq. (1) takes the form
\be
h = h_{ce} + h_q
\ee
where
\be
h_{ce} = JS^2 \sum_{<i,j>} \cos (\theta_i - \theta_j)
\ee
and
$$
h_q = - JS \sum_{<i,j>} \cos (\theta_i - \theta_j) (a^+_i a_i + a^+_j a_j) +
$$
$$
+ {JS\over 2} \sum_{<i,j>} \left[ \cos (\theta_i - \theta_j) + 1 \right] (a^+_i
a_j + a^+_j a_i)
$$
\be
+ {JS\over 2} \sum_{<i,j>} \left[ \cos (\theta_i - \theta_j) - 1 \right] (a_i
a_j + a^+_i a^+_j)
\ee
$$
- {JS\over 2} \sum_{<i,j>} \left[ \cos (\theta_i - \theta_j) - 1 \right]
(\psi^+_i \psi_i + \psi^+_j\psi_j)
$$
$$
+ 2St \sum_{<i,j>}  \cos {\theta_i - \theta_j\over 2}
(\psi^+_i \psi_j + \psi^+_j\psi_i) - \mu \sum_{i} (2S - \psi^+_i\psi_i)
$$

Farther on, we shall use special values for the  angles $\theta_i$
\be
\theta_i = \vec{Q} \cdot \vec{r}_i; \ \ \ \ \ \ Q_x = Q_y = Q
\ee
where $\pi \leq Q \leq 2\pi$ and $\vec{r}_i =(r^x_i, r^y_i)$ are the
coordinates of the lattice site $i$.

The Eqs. (6-8) yield the conventional spin wave theory. To preserve the
sublattice rotational symmetry one must impose, by hand, an additional
condition. Here it is
\be
< \vec{n}_i \cdot \vec{S}_i > = 0
\ee
where $\vec{n}_i$ is a unit vector which fixes the local frame we have chosen
in Eqs.(5).
\be
\vec{n}_i = ( - \sin \theta_i, \ \ 0, \ \ \cos \theta_i)
\ee
When $\theta_i = 2k\pi$ the Eq.(10) is the condition imposed by Takahashi for
ferromagnets$^7$. When $\theta_i = k\pi$ Eq. (10) coincides with the condition
discussed in spin-wave theory of antiferromagnets$^{8,12}$

Making use of Eq. (11) and the representation for the spin operators Eqs. (5),
one can rewrite the condition Eq. (10) in the form
\be
<a^+_i a_i> + {1\over 2} <\psi^+_i \psi_i> = S
\ee
To enforce the constraint Eq. (12) we introduce a new term in the Hamiltonian
\be
h \to h - \lambda \sum_i \left(2 a^+_i a_i + \psi^+_i \psi_i - 2S\right)
\ee
with Lagrange multiplier to be determined by Eq. (12).

It is not difficult to check, that Eq. (12) has a solution just for $Q = \pi ,
2\pi$. This is easy to be understood because the identities
\be
<a_ia_i> = <a^+_i a^+_i> = 0
\ee
are satisfied only for these values of $Q$. To ensure the implementation of the
Eq. (12) for arbitrary $Q$ in the interval $[\pi ,2\pi ]$, one must impose in
addition the condition Eq. (14) also. We introduce in the Hamiltonian a term
with
Lagrange multiplier $\lambda$ which enforces the condition Eq.(12) and a second
term with Lagrange multiplier $\zeta$ which enforces the conditions Eq. (14).
The full Hamiltonian in the momentum space reads
\be
h = 2J S^2 N \cos Q + 2 S N (\lambda - \mu )
\ee
$$
+ \sum_k \left\{ \epsilon_b (k)
a^+_k a_k + D_k [a_k a_{-k} + a^+_k a^+_{-k}]  + \epsilon_f (k) \psi^+_k \psi_k
\right\}
$$
where
$$
\epsilon_b (k) = - 2 \lambda - 4S J \cos Q + 2 S J (1 + \cos Q) \gamma_k
$$
$$
D_k = SJ [\cos Q - 1] \gamma_k + \zeta
$$
$$
\epsilon_f (k) = \mu - \lambda + 2SJ (1-\cos Q) + 8St \cos {1\over 2} Q
\gamma_k
$$
\be
\gamma_k = {1\over 2} (\cos k_x + \cos k_y)
\ee
$N$ is the number of lattice sites, and $\vec{k}$ runs over the Brillouin zone.
Throughout this paper we put the lattice spacing equal to one.

The Bose part of the Hamiltonian Eq. (15) is diagonalized by a Bogoliubov
transformation. The result is
$$
h_b = \sum_k \left[ E_b (k) \alpha^+_k \alpha_k + E_o (k) \right]
$$
where
$$
E_b (k) = \left[ \epsilon^2_b (k) - 4D^2_k \right]^{1/2}
$$
\be
E_o (k)  = {1\over 2} \left[ E_b (k) - \epsilon_b (k) \right]
\ee

The free energy of the system is given by the expression
\be
F = 2JS^2 \cos Q + 2S (\lambda - \mu) + {1\over N} \sum_k E_o (k)
\ee
$$
+{1\over
{\beta N}} \sum_k \ln \left(1 - e^{-\beta E_b(k)}\right) - {1\over
{\beta N}} \sum_k \ln \left(1 + e^{-\beta \epsilon_f(k)}\right)
$$
where $\beta = (k_B T)^{-1}$ is the inverse temperature. The three
chemical potentials $\lambda$, $\zeta$ and $\mu$ are determined by the
equations
$$
{\partial F\over {\partial \lambda}} = 0, \ \
{\partial F\over {\partial \zeta}} = 0, \ \
{\partial F\over {\partial \mu}} = 0
$$
It is more convenient to write their linear combinations
\be
{\partial F\over {\partial \lambda}} + {\partial F\over {\partial \zeta}} =
2S+1 - \delta - {1\over N} \sum_k {\epsilon_b (k) + 2D_k \over{E_b (k)}} {\rm
cth} {\beta \over 2} E_b (k) = 0
\ee
\be
{\partial F\over {\partial \lambda}} - {\partial F\over {\partial \zeta}} =
2S+1 - \delta - {1\over N} \sum_k {\epsilon_b (k) - 2D_k \over{E_b (k)}} {\rm
cth} {\beta \over 2} E_b (k) = 0
\ee
\be
{1\over N} \sum_k {1\over {1 + e^{-\beta \epsilon_f (k)}}} = \delta
\ee
where $\delta$ is the doping parameter.

We  introduce new chemical potentials $\eta_1$, $\eta_2$ which are determined
by the equalities
$$
2S J \eta_1 = -  \lambda - \zeta - 2SJ \cos Q
$$
\be
2S J \eta_2 = -  \lambda + \zeta - 2SJ \cos Q
\ee
Then
$$
\epsilon_b (k) - 2D_k = 4SJ (\eta_1 + \gamma_k)
$$
\be
\epsilon_b (k) + 2D_k = 4SJ (\eta_2 + \cos Q  \gamma_k)
\ee
$$
E_b (k) =  4SJ \sqrt{(\eta_1 + \gamma_k) (\eta_2  + \cos Q  \gamma_k)}
$$

{}From Eqs.(19-23) it follows, that when $Q = \pi , 2\pi$, the solution is
$\eta_1 = \eta_2$ or $\zeta = 0$. This means, that for these special values of
$Q$  the identifies Eq. (14) are satisfied without introducing a second term
in the Hamiltonian and one reobtains the Takahashi's spin-wave theory.

The chemical potentials depend on the temperature and $\eta_1 (T) \geq 1$;\\
$\eta_2 (T) \geq \mid \cos Q \mid$. Next, we analyze the low-temperature
properties of Eqs. (19,20) $\eta_1 (O)$ and $\eta_2 (O)$ depend on $Q$ and
$\delta$. For given doping $\delta$ we shall consider three cases.

When $\pi \leq Q \leq Q_1 \leq {3\over 2} \pi$, we obtain from Eqs. (19,20)
(see Takahashi$^8$)
\be
2S + 1 - \delta - 1.394 \mid \cos Q \mid^{1/2} + {1\over {2\pi SJ \beta}} \ln 2
(4 S J \beta)^2 \mid \cos Q\mid (\eta_1 - 1) = 0
\ee
$$
2S + 1 - \delta - 1.394 \mid \cos Q \mid^{-1/2} + {1\over {2\pi SJ \beta \mid
\cos Q \mid}} \ln 2 (4 S J \beta)^2  (\eta_2 - \mid \cos Q \mid
) = 0
$$

{}From Eq. (24) one gets
$$
\eta_1 - 1 = {1\over {2 (4S J \beta)^2 \mid\cos Q \mid }} e^{-4\pi S J m_1
\beta}
$$
\be
\eta_2 - \mid \cos Q \mid  = {1\over {2 (4S J \beta)^2	}}
e^{-4\pi S J m_2 \mid \cos Q \mid \beta}
\ee
where
$$
2m_1 = 2S + 1 - \delta - 1.394 \mid \cos Q \mid^{1/2}
$$
\be
2m_2 = 2S + 1 - \delta - 1.394 \mid \cos Q \mid^{-1/2}
\ee
$m_1$ increases and $m_2$ decreases as $\mid \cos Q \mid$ decreases, and $m_2 =
0$ for the upper bound of the interval
$$
Q_{(1)} = \arccos \left[ - \left( {1.394\over {2S+1-\delta}}\right)^2\right]
$$
When ${3\pi\over 2} < Q_2 \leq Q \leq 2\pi$ one obtains form Eqs. (19,20)
$$
2S + 1	- \delta - (\cos Q)^{1/2} =
$$
$$
= - {1\over {2\pi S J \beta}} \ln 18S J \Delta_1 {(\Delta_1 + \Delta_2)^{1/2} +
(\Delta_1 + 4\Delta_2)^{1/2} \over{ [2(\Delta_1+\Delta_2)^{1/2} + (\Delta_1 +
4\Delta_2)^{1/2}}}]^2 {\beta\over{(\cos Q)^{1/2}}}
$$
\ \\
$$
2S + 1 - \delta - (\cos Q)^{-1/2} =
$$
\be
= - {1\over{2\pi S J \cos Q \beta}} \ln 2S J [(\Delta_1 + \Delta_2)^{1/2} +
(\Delta_1 + 4\Delta_2)^{1/2}] {\beta\over{ (\cos Q)^{1/2}}}
\ee
where
$$
\Delta_1 = (\eta_1 - 1) (\eta_2 - \cos Q) \cos Q
$$
\be
\Delta_2 = {1\over 4} (\eta_2 - \cos Q \eta_1)^2
\ee
Straightforward calculations lead to the following low-temperature assymptotics
of the chemical potentials $\eta_1$ and $\eta_2$
$$
\eta_1 - 1 = {e^{-M_2 \beta}\over{2SJ(\cos Q)^{1/2} \beta}} \left[ {3+
e^{-(M_1-M_2)\beta} \over{ 9 - e^{-(M_1-M_2)\beta}}} - \left( {1 -
e^{-(M_1-M_2)\beta}\over{ 9 - e^{-(M_1-M_2)\beta}}} \right)^{1/2} \right]
$$
\be
\eta_2 - \cos Q = {(\cos Q)^{1/2}e^{-M_2 \beta}\over{2SJ \beta}}
\left[ {3+ e^{-(M_1-M_2)\beta} \over{ 9 - e^{-(M_1-M_2)\beta}}} + \left( {1 -
e^{-(M_1-M_2)\beta}\over{ 9 - e^{-(M_1-M_2)\beta}}} \right)^{1/2} \right]
\ee
where
\be
M_1 = 4\pi SJm_1; \ \ \ \  M_2 = 4\pi SJm_2 \cos Q
\ee
and $m_1$, $m_2$ are given by
$$
2m_1 = 2S + 1 - \delta - (\cos Q)^{1/2}
$$
\be
2m_2 = 2S + 1 - \delta - (\cos Q)^{-1/2}
\ee
For $Q = 2\pi$, $m_1 = m_2 = m$, $M_1 = M_2$, $\eta_1 = \eta_2 = \eta$ and
\be
\eta - 1 = {1\over{4SJ\beta}} e^{-4\pi Sm}
\ee
For the other values of $Q$, $M_2 < M_1$.  $Q_2$ is defined from Eq. (31) with
$m_2= 0$
\be
Q_2 = \arccos {1\over {(2S + 1 - \delta)^2}}
\ee

Finally, let $Q$ runs the interval $Q_1 < Q < Q_2$. Then $\eta_1 (T)$ goes to
one, when $T$ goes to zero, and $\eta_2 (O) > \mid \cos Q \mid$.
\be
\eta_1 - 1 = {1\over {(4SJ\beta)^2 (\eta_2 - \cos Q)}} e^{-4\pi SJ m_1 \beta}
\ee
where $m_1$ and $\eta_2(O)$ are determined by the equations
$$
2S + 1 - \delta = {1\over N} \sum_k \left( {\eta_2 + \cos Q \gamma_k \over{ 1 +
\gamma_k}}\right)^{1/2} + 2m_1
$$
\be
2S + 1 - \delta = {1\over N} \sum_k \left( {1 + \gamma_k  \over{
\eta_2 + \cos Q \gamma_k}}\right)^{1/2}
\ee
and $m_2 = 0$.

Let us recapitulate the results we have received above. When the temperature
goes to zero $\eta_1 (T)$ goes to one for all values of $Q$. At the same time
$\eta_2(O) = \mid \cos Q \mid$ if $\pi \leq Q \leq Q_1$, and $m_1,m_2$ can be
found from Eqs.(26). When $Q_1 < Q < Q_2$, $m_2 = 0$ and $\eta_2 (O)$ and $m_1$
can be obtained from the system Eqs.(35). Finally, when $Q_2 \leq Q \leq 2\pi$,
$\eta_2 = \cos Q$ and $m_1, m_2$ are given by Eqs. (31).

For the further convenience we have used a common notation for the variables
$m_1, m_2$ when $\pi \leq Q \leq Q_1$ and when $Q_2 \leq Q \leq 2\pi$, but
their origin is different. At zero temperature, when $\pi \leq Q \leq Q_1$,
$\eta_1 = 1$ and $\eta_2 = \mid \cos Q \mid$, the bose system condenses at the
wave vectors $k^* = (\pm \pi,  \pm \pi)$ and $k_o = (0,0)$. Then, the spin-wave
correlators read
$$
<a^+_ka_k> = {1\over 2} \left[ {\epsilon_b (k)\over{E_b(k)}} {\rm cth}
{\beta \over 2} E_b (k) - 1 \right] (1 - \delta_{k,k^*}) \left( 1 -
\delta_{k.k_o}\right)  $$
\be
+ {1\over 2} n_{k^*} \delta_{k,k^*} + {1\over 2} n_{k_o} \delta_{k,k_o}
\ee
\ \\
$$
<a_k a_{-k}> = <a^+_k a^+_{-k}> =
$$
$$
= -{D_k \over{ E_b(k)}} {\rm cth} {\beta\over 2} E_b (k) (1 - \delta_{k,k^*})
(1 - \delta_{k,k_o}) - {1\over 2} n_{k^*} \delta_{k,k^*} + {1\over 2} n_{k_o}
\delta_{k,k_o}
$$
and
\be
m_1 = {1\over N}  \sum_{k^*} n_{k^*}, \ \ \ \ m_2 = {n_{k_o}\over N}
\ee
When $Q_2 \leq Q \leq 2\pi$, $\eta_1 = 1$ and $\eta_2 = \cos Q$ the bose system
condenses at the wave-vector $k^*$ and for the spin-wave correlators are
obtains
$$
<a^+_k a_k> = {1\over 2} \left[ {\epsilon_b (k)\over{E_b(k)}} {\rm cth}
{\beta \over 2} E_b (k) - 1 \right] \left( 1 - \delta_{k,k^*}\right)  + {1\over
2} n^{(1)}_{k^*} \delta_{k,k^*}
$$
\be
<a_k a_{-k}> = <a^+_k a^+_{-k}> = - {D_k\over{E_b(k)}} {\rm cth}
{\beta \over 2} E_b (k)\left( 1 - \delta_{k.k^*}\right) - {1\over
2} n^{(2)}_{k^*} \delta_{k,k^*}
\ee
where
$$
{1\over {2N}} \sum_{k^*} (n^{(1)}_{k^*} + n^{(2)}_{k^*}) = m_1
$$
\be
{1\over {2N}} \sum_{k^*} (n^{(1)}_{k^*} - n^{(2)}_{k^*}) = m_2
\ee
When $Q_1 < Q < Q_2$ the spin-wave correlators are given by Eq.(38) with
$n^{(1)}_{k^*} = n^{(2)}_{k^*}$ ($m_2 = 0$).

To complete our analysis we shall discuss the magnon spectrum as a function
of $Q$.  From Eqs (23) we obtain that when $\pi \leq Q \leq Q_1$ there are two
antiferromagnetic magnons: $E_b(k) \simeq v_s \mid k - k^* \mid$ when
$\mid k - k^* \mid \to 0$ and $E_b (k) \simeq v_s \mid k - k^o \mid $
when $\mid k - k^o \mid \to 0$, with spin-wave velocity $v_s = 4 S J
\sqrt{\mid \cos Q \mid /2}$. When $Q$ exceeds $Q_1$ one of the magnons opens a
gap $E_b \simeq v_s \sqrt{\Delta^2 + \mid k - k^0 \mid^2}$ where
$\Delta^2 = 4 (\eta_2 - \mid \cos Q \mid) / \mid \cos Q \mid$. In the interval
$(3\pi /2, Q_2)$ there is only one magnon
$E_b \simeq v'_s \mid k-k^* \mid$ with
$v'_s =  2 S J (\eta_2 - \cos Q)^{1/2}$. The spin-wave velocity
$v'_s$  approaches zero when $Q$ approaches $Q_2$ and in the interval
$[Q_2, 2\pi]$ we obtain the ferromagnetic magnon $E_b \simeq S J \sqrt{\cos Q}
\mid k - k^* \mid^2$.

The angle $Q$ is a free parameter in the theory. To determine it we shall
calculate the energy of the system per site.

{}From Eqs.(5), taking into account the conditions Eqs.(12,14), one obtains in
spin-wave approximation
$$
<\vec{S}_i \cdot \vec{S}_j> =
$$
\be
\cos (\theta_i - \theta_j) \left[ <a^+_i a_j> <a^+_j a_i> + <a^+_ia^+_j>
<a_ia_j> - {1\over 4} <\psi^+_i\psi_j><\psi^+_j\psi_i> \right]
\ee

$$
<c^+_{i\sigma}c_{j\sigma} + h.c.> =
$$
\be
= - \cos {\theta_i - \theta_j \over 2} \left[ (S + {1\over 2} \delta)
\left(<\psi^+_j\psi_i> + <\psi^+_i\psi_j> \right) + \right.
\ee
$$
\left . +<\psi^+_j\psi_i><a^+_ia_j> + <\psi^+_i\psi_j><a^+_ja_i> \right]
$$
and
$$
<n_in_j> = (2S - \delta)^2 - <\psi^+_i\psi_j><\psi^+_j\psi_i>
$$

Putting Eqs.(40-42) into Eq.(1) one obtains for the energy per site at $T=0$
and $\pi \leq Q \leq Q_2$
 $$
E = -{J\over 2} (2S - \delta)^2 + {1\over 2} (1 - \cos Q) \left[ {1\over N}
\sum_k \gamma_k \theta \left(-\epsilon_f (k) \right) \right]^2 +
$$
$$
+{J\cos Q\over 4} \left[ 2S + 1 - \delta -
{1\over N} \sum_k (1 - \gamma_k) \left( {1+\gamma_k\over {\eta_2 + \cos Q
\gamma_k}} \right)^{1/2} \right]^2 +
$$
$$
+ {J\cos Q\over 4} \left[ 2S + 1 - \delta -
{1\over N} \sum_k (1 + \gamma_k) \left( {\eta_2 + \cos Q
\gamma_k\over {1+\gamma_k }} \right)^{1/2} \right]^2 +
$$
$$
+2t \cos {Q\over 2} \left[ {1\over N} \sum_k \gamma_k \theta
\left(-\epsilon_f(k) \right)\right]\left[ {1\over {2N}} \sum_k (1 + \gamma_k)
\left( {\eta_2 + \cos Q \gamma_k\over {  1+\gamma_k }} \right)^{1/2} -
\right .
$$
\be
\left .   -  {1\over {2N}} \sum_k (1 -
\gamma_k) \left( {1+\gamma_k\over {\eta_2 + \cos Q
\gamma_k}} \right)^{1/2} + 2S + \delta \right]
\ee
where $\eta_2$ is defined above and $\theta(-\epsilon_f)$ is the step
function.

When $Q_2 \leq Q \leq 2\pi$, the energy is given by
$$
E = - {J \over 2} (2S - \delta)^2 +  {1\over 2} (1 - \cos Q) \left[ {1\over N}
\sum_k \gamma_k \theta \left( -\epsilon_f (k) \right) \right]^2
$$
$$
+{J \cos Q\over 4} \left[ 2S + 1 - \delta - {1 \over { \sqrt{ \cos Q}}}
\right]^2  +
$$
$$
+{J \cos Q\over 4} \left[ 2S + 1 - \delta -  \sqrt{ \cos Q}\right]^2 +
$$
$$
+ 2t \cos {Q\over 2} \left[ {1\over N} \sum_k \gamma_k \theta \left(
-\epsilon_f (k) \right) \right] \left[ 2\delta - 1 + {1\over {2\sqrt{\cos Q}}}
+ {\sqrt{\cos Q} \over 2} \right]
$$

For a given $\delta$ $E$ is a function of $Q$, and the physical value of $Q
(Q_o)$ is that which minimizes the energy. It depends on $\delta$.

When the doping is very small $Q_0$ is close to $\pi$, $\eta_2 = \mid \cos Q
\mid$, and
$$
{1\over N} \sum_k \gamma_k \theta (-\epsilon_f (k)) \simeq \delta
$$

The minimum of the energy is at
\be
Q_o = \pi + {S\over {(S+0.079)^2}} {t\over J} \delta
\ee

The energy per site is plotted in Fig.1 as a function of $Q$, for different
values of $\delta$, $t/J = 5$ and $S = 1/2$. When $\delta$ increases the
minimum of the energy moves smoothly. A second minimum at $Q = 2\pi$ appears
when the doping is large enough. This minimum decreases when $\delta$ increases
and at some value of the doping the both minimums become equal. At this point
$Q_o$ has a step-wise behavior. The curves $Q_o (\delta)$ are plotted in Fig.
2 for different values of $t/J$ and $S = 1/2$. We treat the appearance of the
second minimum as a appearance of instability and as a upper bound of
applicability of our approach. For $t/J = 10$ this bound is $\delta = 0.31$,
for $t/J = 5$; $\delta = 0.39$, for $t/J = 3$; $\delta = 0.47$ and for $t/J =
0,5$; $\delta = 0.6$.
\ \\

{\bf III. The spin-spin correlator}

Let us consider the spin-spin correlation function in the limit of large
separation between the two sites. At zero temperature the bose system
condenses. Then, the large distance asymptotic of the bose correlators is
determined by the values of their Fourier transforms at the wave vectors at
which the system condenses. From Eqs.(36-39) one obtains
\be
<a^+_ia_j><a^+_ja_i> + <a_ia_j><a^+_ia^+_j> \simeq m^2
\ee
where
\be
m^2 = {1\over 2} (m_1^2 +  m^2_2)
\ee

 The fermi term goes to zero when $\mid r_i - r_j \mid$ goes to infinity and
we shall drop it.  As a result we  obtain that the large
distance asymptotic of spin-spin correlator at $T=0$  is
\be
<\vec{S}_i \cdot \vec{S}_j> = m^2 \cos Q_o \cdot (r_i - r_j)
\ee

This  implies the presence of the long-range spiral phase.
{}From Eqs.(26) one obtains that in leading order of $\delta$
\be
m_1 = m_2 = m = S - 0.197 - {1\over 2} \delta
\ee

In Fig. 3 we have plotted $m$ as a function of $\delta$ for different values of
$t/J$.

At finite temperature, the bose energy $E_b(k)$ (see Eq. (23)) has a finite
gap. As a consequence, the bose correlators have a finite correlation length.
For the spin-spin correlation function at small doping (when $\pi \leq Q_o
(\delta) \leq Q_2 (\delta)$    and large distance one obtains
$$
<\vec{S}_i \cdot \vec{S}_j> \simeq {1\over {(2S J_{\beta})^2}} {\cos Q_o \cdot
(r_i - r_j) \over {2\pi \mid r_i -  r_j \mid}} \left[ \xi_1 e^{-{\mid r_i - r_j
\mid \over {\xi_1}}} + \right .
$$
\be
\left . + {\xi_2 \over {(\cos Q_o)^2}}	e^{-{\mid r_i - r_j
\mid \over {\xi_2}}} \right]
\ee
where
\be
\xi^{-1}_1 = 4 \sqrt{\eta_1 - 1}
\ee
\be
\xi^{-1}_2 = 4 \sqrt{(\eta_2 - \mid \cos Q_o \mid ) / \mid \cos Q_o \mid }
\ee

{}From Eqs. (25) it follows that $\xi_1$ and $\xi_2$ grow exponentially when
$T$
goes to zero. At very small doping $\eta_1 \simeq \eta_2$, $\xi_1 \simeq
\xi_2$, $\mid \cos Q_o \mid \simeq 1$ and the spin-spin correlator looks like
in the theory of the Heisenberg antiferromagnet$^8$. When $\delta$ increases
$m_2$ decreases and the first term in Eq.(48) is dominated when $T$ approaches
zero.
\ \\

{\bf IV. Concluding remarks}

 In conclusion, we have investigated the two-dimensional $t-J$ model by
modified spin-wave theory. A spin-spin correlator, we have obtained, at $T = 0$
 Eq. (46) shows a spiral long-range order. At $T\neq 0$
the correlation function decays exponentially as in the theory of the
Heisenberg
antiferromagnet. The results, at small doping,	are in	qualitative
agreement with the results
obtained by Schwinger boson slave-fermion  mean field theory$^5$, except for
the vector 3/2 known from the earlier publications. Substantial quantitative
discrepancies appear when $\delta$ increases. In spite of this the approach we
have discussed above seams to be appropriate for the investigation of the
normal state of the novel superconductors.

As a by-product we obtain the spin-stiffness constant $\rho_S$ of the square
lattice Heisenberg antiferromagnets. It can be defined by
$$
\rho_S = \left.{d^2E\over{dQ^2}} \right|_{Q = \pi}
$$
where $E(Q)$ is the energy per site of the system. From Eq. (42) at $\delta
= 0$ one obtains
$$
\rho_S = 2J (S + 0.079)^2
$$
{\bf References} \\
\begin{enumerate}
\item P.W.Anderson, Science, \underline{235}, 1196 (1987).
\item C.Gross, R.Joynt, and T.M.Rice, Phys.Rev. \underline{B36}, 381
(1987).
\item F.Zhang and T.M.Rice, Phys.Rev. \underline{B37}, 3759 (1988).
\item D.Arovas and A.Auerbach, Phys.Rev. \underline{B38}, 316 (1988);\\
D.Yoshioka, J.Phys.Soc.Jpn. \underline{58}, 32 (1989).
\item D.Yoshioka, J.Phys.Soc.Jpn. \underline{58}, 1516 (1989); \\
C.Jayaprahash, H.R.Krishnamurty, and S.Sarker, Phys.Rev. \underline{B40}, 2610
(1989); \\
K.Feisenberg, P.Hedeg\aa rd, and M.Pederson, Phys.Rev. \underline{B40}, 850
(1989).
\item  B.Shraiman and E.Siggia, Phys.Rev.Lett \underline{62}, 1564 (1989).
\item M.Takahashi, Prog.Theor.Phys. Suppl. \underline{87}, 233 (1986);
Phys.Rev.Lett. \underline{58}, 168 (1987).
\item M.Takahashi, Phys.Rev. \underline{B40}, 2494 (1989).
\item S.B.Chakravarty, B.I.Halperin, and D.Nelson, Phys.Rev.
\underline{B39}, 2344 (1989).
\item N.I.Karchev, Theor. and Math. Phys. October 1992.
\item A.Angellucci and R.Link, MIT preprint, CTP$\sharp$ 2043 (1991).
\item J.E.Hirsh and S.Tang, Phys.Rev. \underline{B40}, 4769 (1989).
\end{enumerate}
\newpage

\begin{center}FIGURE CAPTIONS \end{center}

Fig. 1. The energy of the system per site as a function of $Q$ for $t/J = 5$,
$S = 1/2$ and different values of doping.
\ \\

Fig. 2. The physical value $Q_o$ of the spirality angle as a function of
doping, for $t/J = 10$ (solid line with circles), $t/J = 5$ (dashed line), $t/J
= 3$ (dotted line), $t/J = 0.5$ (dot-dashed line) and $S = 1/2$.
\ \\

Fig. 3. The quantity $m$ as a function of doping for $t/J = 10$ (solid line),
$t/J = 5$ (dashed line), $t/J = 3$ (dotted line), $t/J = 0.5$ (dot-dashed line)
and $S = 1/2$.

\end{document}